\documentclass[12pt]{article}

\RequirePackage{etoolbox}
\usepackage{navigator}

\newenvironment{wileykeywords}{\textsf{Keywords:}\hspace{\stretch{1}}}{\hspace{\stretch{1}}\rule{1ex}{1ex}}

\usepackage{amsmath,amssymb}
\usepackage{graphicx}
\usepackage{color}
\usepackage{dcolumn}
\usepackage{bm}
\usepackage[numbers,super,comma,sort&compress]{natbib}
\usepackage{upgreek}
\usepackage{multirow}

\newcommand{\citeonl}[1]{[{}$\!\!$\citenum{#1}]}
\newcommand{\citeonlb}[1]{[{}$\!\!\!$\citenum{#1}]}

\usepackage[pdfstartview=FitH,bookmarksopen=true]{hyperref}

\definecolor{background-color}{gray}{0.98}

\title{\sffamily Machine learning for accuracy in density functional approximations}
\author{%
Johannes Voss\thanks{SUNCAT Center for Interface Science and Catalysis, SLAC National Accelerator Laboratory, 2575 Sand Hill Road, Menlo Park, CA 94025, USA}
\thanks{vossj@slac.stanford.edu\medskip\\
This is the peer reviewed version of the following article: Voss.~J.\ {\it J.~Comput.~Chem.}\ 45, 1829--1845 (2024), which has been published in final form at https://doi.org/10.1002/jcc.27366. This article may be used for non-commercial purposes in accordance with Wiley Terms and Conditions for Use of Self-Archived Versions. This article may not be enhanced, enriched or otherwise transformed into a derivative work, without express permission from Wiley or by statutory rights under applicable legislation. Copyright notices must not be removed, obscured or modified. The article must be linked to Wiley’s version of record on Wiley Online Library and any embedding, framing or otherwise making available the article or pages thereof by third parties from platforms, services and websites other than Wiley Online Library must be prohibited.}
}
\date{October 2, 2025}

\begin{document}

\maketitle

\addcontentsline{toc}{section}{\sffamily \large ABSTRACT}

\begin{abstract}
Machine learning techniques have found their way into computational chemistry as indispensable tools to accelerate atomistic simulations and materials design. In addition, machine learning approaches hold the potential to boost the predictive power of computationally efficient electronic structure methods, such as density functional theory, to chemical accuracy and to correct for fundamental errors in density functional approaches. Here, recent progress in applying machine learning to improve the accuracy of density functional and related approximations is reviewed. Promises and challenges in devising machine learning models transferable between different chemistries and materials classes are discussed with the help of examples applying promising models to systems far outside their training sets. 
\end{abstract}

\begin{wileykeywords}
Machine learning, density functional theory, materials prediction, exchange-correlation functional, self-interaction, electron delocalization
\end{wileykeywords}

\clearpage


\begin{figure}[h]
\centering
\colorbox{background-color}{
\fbox{
\begin{minipage}{1.0\textwidth}
\includegraphics[width=50mm,height=50mm]{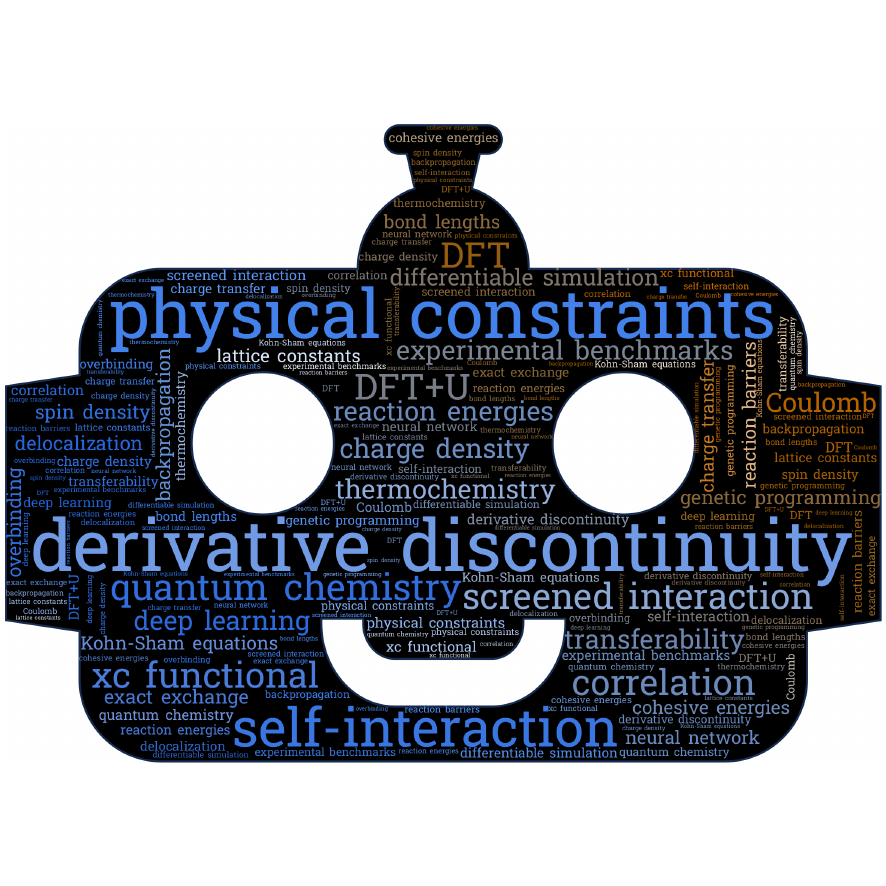} 
\\
Machine learning techniques allow us, where benchmark data are available, to train electronic structure models that substantially increase the predictive power of density functional theory simulations of chemical reactions and structural and thermodynamic properties of gas, liquid, and solid phases. Not only can quantitative improvements be achieved, but also fundamental limitations of density functional approximations can be corrected for. Here, techniques, benchmark data, and challenges for devising transferable electronic structure machine learning models are reviewed.
\end{minipage}
}}
\end{figure}

  \makeatletter
  \renewcommand\@biblabel[1]{#1.}
  \makeatother

\bibliographystyle{chem}

\renewcommand{\baselinestretch}{1.15}
\normalsize

\clearpage

\section{\sffamily \large INTRODUCTION}

Machine learning (ML) techniques play an increasingly important role in atomistic-scale simulations in computational chemistry and physics.\cite{Keith2021,Schmidt2019,Retrospective2020} Major areas of research are the acceleration of materials discovery and extending computationally accessible time and length scales through accelerated simulations. Inter-atomic potentials represented by neural networks\cite{Behler2007,schnet,Smith_ani2017,aenet} or other ML regression techniques\cite{GAP2010,Chmiela2017} enable accurate molecular dynamics simulations for system sizes and time scales well beyond what can be achieved with first-principles Hamiltonians.\cite{Unke2021} When computation of the Born-Oppenheimer potential energy surface is not required, ML approaches trained to map chemical composition and other not necessarily atomic structure sensitive features to system properties of interest are powerful methods for direct, approximate materials property predictions.\cite{Isayev2017,Xie2018,Hautier2010,NIPS2015,Ward2016} Such methods can furthermore be employed for inverse materials design, where molecules or materials compositions that could lead to a desired target metric are predicted.\cite{Kim2018,Noh2019} These models and inter-atomic potentials are trained on high-throughput datasets generated with computationally affordable methods. Density functional theory (DFT)\cite{hohenberg1964inhomogeneous} is often the method of choice due to a favorable trade-off between computational complexity and accuracy for the prediction of the electronic structures of molecules and solids.\cite{Kohn1996,BurkeDFT} Some (minor or appreciable) loss in accuracy with respect to the DFT training data is typically tolerated with the advantage of significant speed up of the resulting ML methods over DFT simulations. At best, these ML methods can reproduce the quality of the DFT training data.

\begin{figure}
\centering
\includegraphics[width=3.33in,keepaspectratio=true]{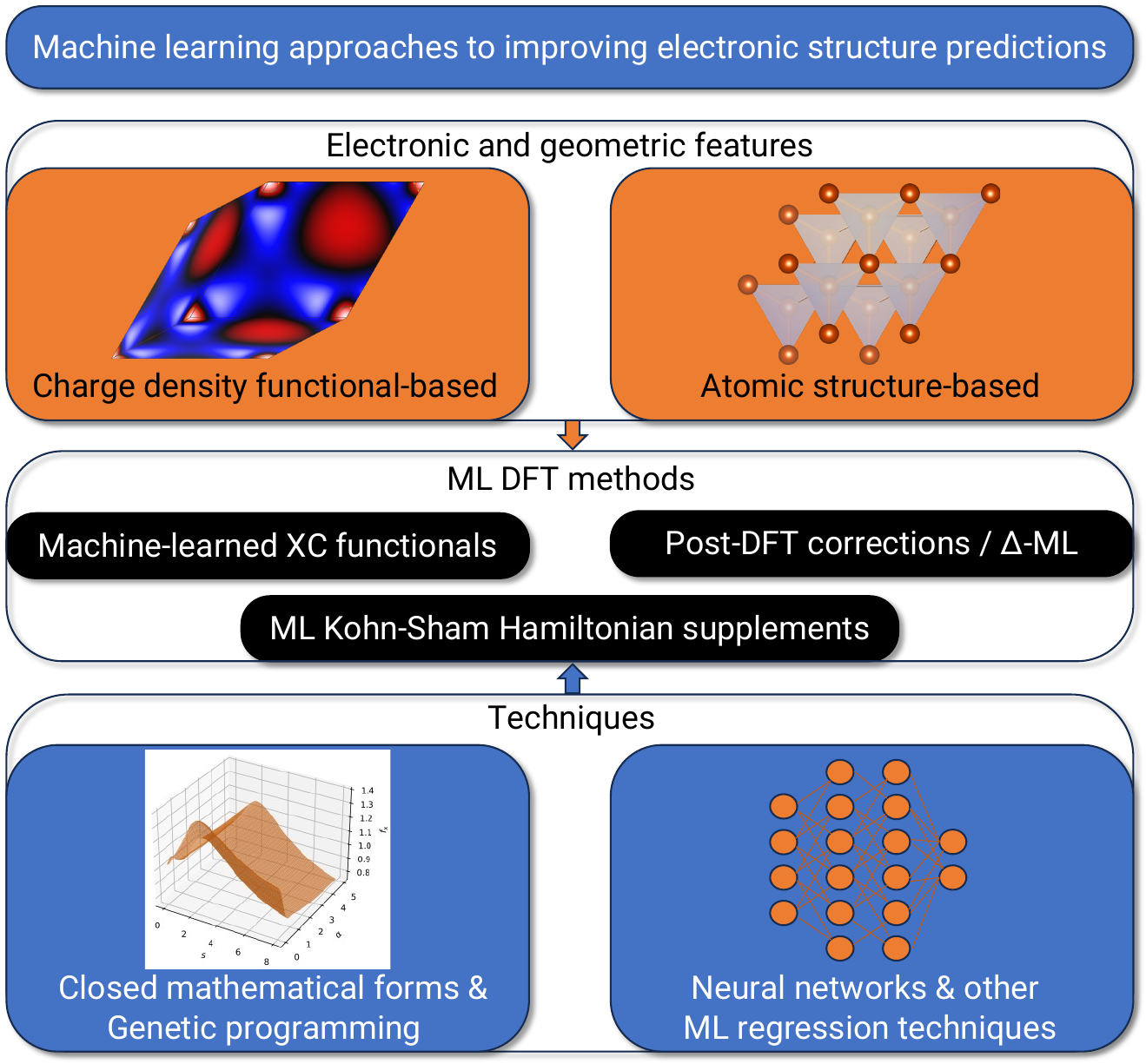}
\caption{\label{fig:overview} Overview of ML approaches to increasing the accuracy of electronic structure predictions based on electronic or atomic structural features. Machine-learned XC functionals are trained on high-accuracy benchmark data to improve upon the predictive power of existing DFAs. Post-DFT and $\Delta$-ML methods provide improved energetics on fixed DFT charge densities, and other ML approaches supplement the Kohn-Sham Hamiltonian with Hubbard and dispersion terms.}
\end{figure}

ML approaches are, however, also employed to increase the accuracy of DFT and related methods rather than substituting these first-principles approaches completely with ML models for acceleration. The ML methods are trained against chemically accurate quantum chemistry reference data or experimental benchmark data, where sufficient accuracy with beyond-DFT methods currently cannot be achieved. These methods can be categorized (Figure~\ref{fig:overview}) into machine-learned density functionals for exchange and correlation (XC), atomic structure-dependent, machine-learned Hamiltonian corrections, and $\Delta$-ML approaches that learn a correction to be applied to DFT results (as post-DFT corrections), with some methods belonging to more than one of these categories. Here, recent progress in ML approaches to increasing accuracy and to correcting fundamental errors in density functional approximations (DFA) is reviewed. These approaches hold the promise of providing DFT predictions with chemical accuracy and enabling accurate electronic structure simulations where DFAs fundamentally fail and which are currently out of reach for higher levels of theory. There are, however, challenges in availability of accurate training data for these latter systems and there can be issues with transferability of the ML methods beyond their training data. Examples are provided demonstrating such transferability issues for promising ML models.

\section{\sffamily \large SHORTCOMINGS OF DENSITY FUNCTIONAL APPROXIMATIONS}

DFT drastically reduces the complexity of the electronic structure problem by expressing the total energy $E_{\rm total}$ of a system as a functional of the electronic charge density $\rho$ rather than the many-electron wave function. In the Kohn-Sham (KS) formulation of DFT, this density functional is
\begin{equation}
E_{\rm total}[\rho] = T_{\rm KS}[\rho] + E_{\rm XC}[\rho] + E_{\rm H}[\rho] + E_{\rm ext}[\rho] .\label{eq:kohnsham}
\end{equation}
$T_{\rm KS}[\rho]$ is the kinetic energy of an auxiliary system of non-interacting particles with the same density $\rho$ as the true system of interacting electrons. Functional differentiation $\delta E_{\rm total}/\delta \rho$ leads to a set of single-particle like KS equations with an effective one-body KS potential, from which $T_{\rm KS}[\rho]$ is computed. $E_{\rm ext}[\rho]$ accounts for the interaction with an external potential (given, {\it e.g.}, by interaction with nuclei) and $E_{\rm H}[\rho]$ for the electrostatic interaction of the density with itself. The XC functional $E_{\rm XC}[\rho]$ accounts for the two-body Coulomb interaction of the electrons and corrects for self-interaction in $E_{\rm H}[\rho]$ and differences between $T_{\rm KS}[\rho]$ and the kinetic energy of interacting electrons. In principle, Eq.~\ref{eq:kohnsham} is exact, was $E_{\rm XC}[\rho]$ known. In practice, $E_{\rm XC}[\rho]$ must be approximated.

A few of the issues of approximations to $E_{\rm XC}[\rho]$, some of which could be improved upon using ML methods, are summarized below. For a review of the limitations of DFAs, the reader is referred to Refs.~\citeonl{PerdewJCTC2008,Cohen2012,Becke2014,Verma2020,Bryenton2023}.

The exact (unknown) $E_{\rm XC}[\rho]$ is a universal functional not depending on the system in consideration. Approximations to it typically perform better for prediction of some materials properties at the cost of a worse prediction of others. Generalized gradient approximations (GGA), in which $E_{\rm XC}[\rho]$ is expressed locally as a functional of the density and its gradient, improve upon the simplest approximation only depending on the local density (local density approximation: LDA; local spin density approximation: LSDA), which is fitted accurately\cite{PerdewWang1992,PerdewZunger1981} against Quantum Monte Carlo (QMC) simulations of the homogeneous electron liquid.\cite{CeperleyAlder} Some GGAs perform relatively well for solid structural and elastic properties but not for reaction energetics and vice versa.\cite{pbesol,wucohen,revpbe,rpbe} Meta-GGAs, that additionally dependent on the Kohn-Sham kinetic energy density $\tau$ or the Laplacian of $\rho$ improve the range of applicability over GGA approaches.\cite{TPSS,SCAN,REVM06L,RTPSS,MCML,OFR2}

The spurious electrostatic interaction of an electron with itself contained in the Hartree term $E_{\rm H}$ is difficult to compensate with the above semi-local functionals in strongly inhomogeneous systems or systems with localized states due to the long range of Coulomb interactions. One consequence of this inherent difficulty is spurious charge transfer across inter atomic separations at which the interaction between the separated subsystems should have vanished leading to neutral atoms. Hartree-Fock (HF) exchange exactly cancels this type of self-interaction error, and hybrid functionals combine semi-local DFT with (some amount of) such exact exchange (EXX) energies (at increased computational cost over semi-local DFT).\cite{Becke} Range-separated hybrids consider screened exchange integrals, retaining EXX only at short or long range. Long-range EXX\cite{Iikura2001,Leininger1997} is required to cancel the long-range Hartree self-interaction. Metallic systems, on the other hand, are incorrectly described by long-range EXX with a vanishing density of states at the Fermi level. Metals are thus studied with short-range hybrids,\cite{hse03} which unfortunately do not cancel the long-range part of the Hartree self-interaction. These conflicting requirements for short {\it vs.}\ long-range EXX make it particularly difficult to accurately model the interaction of molecules with metallic surfaces with hybrid DFT.\cite{ADS41} One-electron self-interaction errors can generally be addressed with the self-interaction correction scheme by Perdew and Zunger,\cite{PerdewZunger1981} improving, {\it e.g.}, charge transfer energetics and the description of negative ions. The prediction of thermochemistry is, however, worsened over uncorrected DFAs,\cite{Vydrov2004} and also equilibrium geometries of molecules and solids are not systematically improved.\cite{Perdew2015}

Even for gapped systems that can be studied with long-range hybrids, fundamental problems in the predicted electronic structures can exist in the presence of strong static correlation. Since the auxiliary KS system is composed of non-interacting particles, $T_{\rm KS}$ is computed from a single Slater determinant of KS orbitals. While the exact $E_{\rm XC}[\rho]$ could compensate for the difference to the true, interacting electronic kinetic energy with significant multi-determinantal contributions (unless the true ground state density should turn out to not be representable by a non-interacting system with local KS effective potential\cite{PerdewJCTC2008}), approximations to $E_{\rm XC}[\rho]$ typically lead to significant qualitative errors in {\it e.g.}\ predicting the energetics of multiradical molecules.\cite{Mayhall} While KS spin-orbital occupation-constrained DFT can be used to compute corrections to some of these static correlation errors,\cite{Noodleman,Daul1994} in the general case of molecular interactions or molecule-metal interaction, where such constraints cannot be straightforwardly or uniquely defined or applied, static correlation is a fundamental problem for DFT and hybrid DFT. Long-range hybrids plus orbital-dependent random phase approximation\cite{PhysRev.82.625} (RPA) correlation were shown to improve upon the description of strong static correlation in the dissociation limit of singly positively charged dimers.\cite{Paier2010}

Another shortcoming of DFAs is the incorrect prediction of total energies as a function of number of electrons $N$. Fractional values of $N$ correspond to quantum mechanical ensemble averages of systems with different integer electron counts. Between adjacent integer values of $N$, the total energy scales linearly with $N$ with derivative discontinuities at integer $N$.\cite{DerivDiscont} Semi-local DFAs do not reproduce the linear behavior nor the derivative discontinuities but rather show a convex behavior of $E_{\rm total}(N)$, thus predicting too low energies for fractional $N$ corresponding to a delocalization error favoring overly delocalized charge distributions over more localized ones with integer occupations.\cite{MoriSanchez2006} Supplementing the KS Hamiltonian with Hubbard terms in DFT+$U$ approaches\cite{AnisimovPRB1991,CzyzykPRB1994,LiechtensteinPRB1995} canceling\cite{CococcioniPRB2005} the spurious curvature of $E_{\rm total}(N)$ fundamentally improves the description of systems with strong $d$-electron or $f$-electron localization, such as several insulating transition-metal oxides and lanthanides and actinides, respectively. Predicting chemical reaction energies with DFT+$U$ approaches is, however, difficult, as only total energies with same Hubbard $U$-parameters for each ion in products {\it vs.}\ reactants can be compared directly. Transition-metal oxide formation energies computed with average, empirical $U$-parameters applied both to oxides and metals are an improvement over GGA predictions,\cite{WangPRB2006} at the prize of a worse description of the bandstructure of metallic phases.\cite{Voss2022} The lack of the derivative discontinuity and resulting delocalization error for semi-local DFAs furthermore can be used to explain the DFA band gap problem of generally underestimating the band gaps of semiconductors and insulators.\cite{MoriSanchezPRL2008} Orbital-dependent DFAs were constructed to provide derivative discontinuity corrections that added to KS band gaps approximately yield the desired fundamental gaps.\cite{Kuisma2010}

Semi-local DFAs fail at describing van der Waals (vdW) interactions. Here, density-density convolution approaches\cite{Andersson1996,PhysRevLett.92.246401,Vydrov2010} provide nonlocal DFA corrections which are quite successful in describing dispersion forces. Alternatively, force fields (with fixed or density-dependent dispersion coefficients) can correct the potential energy surfaces computed with semi-local DFAs by vdW interactions.\cite{GrimmeD2,GrimmeD3,GrimmeD4,Tkatchenko2009,Sato2009,Steinmann2011}

For molecular systems of sufficiently low electron count $N$, there are wave function-based quantum chemistry methods, which can provide accurate electronic structures, not suffering from the above problems of DFAs. For extended systems, in particular those involving metallic states, obtaining results more accurate than those from semi-local DFAs is difficult, and experiments often serve as the benchmark for DFAs. In the following section, reference data that can serve for training electronic structure ML models are summarized.

\section{\sffamily \large GROUND TRUTH FOR DFA ML MODELS}

\subsection{Thermochemistry, thermochemical kinetics, and molecular interactions}

Experimental data for heats of formation of molecules are common benchmarks for DFAs and quantum chemistry ({\it i.e.}\ generally wave function-based) methods. The ``Gaussian-$n$'' theories of composite quantum chemistry techniques,\cite{Gaussian3,Gaussian4} {\it e.g.}, were benchmarked against experimental data on heats of formation, ionization potentials, electron affinities, and proton affinities, with $\sim$1~kcal/mol errors.\cite{G399,G305} Having thus established chemical accuracy of the approaches for these molecular properties, further accurate training and benchmark data for DFAs and ML models can be computed in high throughput with these quantum chemistry methods. Ramakrishnan {\it et al.}\cite{Ramakrishnan2014} used the Gaussian-4-M{\o}ller-Plesset-2 level of theory\cite{Gaussian4reduced} to compute the heats of formation of over 100,000 molecules constituting the QM9 dataset (out of $\sim2\cdot10^{11}$ molecules enumerated in the GDB-17 dataset\cite{gdb17}). Using the Weizmann-4 quantum chemistry protocol,\cite{weizmann1,weizmann2} Karton {\it et al.}\ were able to compute training targets for atomization energies of 140 molecules and radicals in the W4-11 dataset\cite{Karton2011} and of 200 molecules and radicals in the W4-17 dataset\cite{Karton2017} with estimated accuracies better than 1~kcal/mol, also providing computed zero-point energies.

Řezáč {\it et al.}\cite{s66,s66ext} computed the non-covalent interaction energies of 66 molecular complexes at 8 inter-molecular separations using coupled cluster theory\cite{coupledclusters} with triple excitations and complete basis set limit extrapolation. These S66/S66x8 datasets (later revised by Brauer {\it et al.}\cite{Brauer2016}) are particularly useful for parameterizing or training approaches to describe dispersion energetics and forces.

Gas phase reaction barrier height benchmarks were established by combining several quantum chemistry approaches and experimental results by Zheng {\it et al.}\cite{dbh,dbh2} in the DBH24 dataset and by Zhao {\it et al.}\cite{zhao2004development,zhao2005benchmark} in the BH-76 dataset. These two barrier height and other datasets are included in the GMTKN55 dataset by Goerigk {\it et al.},\cite{Goerigk2017} which is a collection of datasets for thermochemistry, thermochemical kinetics, and non-covalent interactions. A further aggregate molecular chemistry benchmark dataset was compiled by Mardirossian and Head-Gordon.\cite{Mardirossian2017} Chan {\it et al.}\cite{Chan2019} collected quantum chemistry benchmark data for transition-metal (complex) chemistry in the TMC151 dataset.

For heats of formation of solids, the tables of Kubaschewski {\it et al.}\cite{Kubaschewski1993} provide a wide range of experimental data. These data were {\it e.g.}\ used to parameterize empirical schemes to combine DFT+$U$ energetics for correlated transition-metal oxides with DFT for metallic phases in the Materials Project.\cite{JainPRB2011,JainAPLM2013} Kirklin {\it et al.}\cite{OQMD} have collected experimental solid formation energies for benchmarking their Open Quantum Materials Database of GGA simulations of solids. For solid cohesive energies, experimental data collected in Refs.~\citeonl{mbeefvdw,ZhangNJP2018} was combined in the CE65 dataset, where zero-point contributions estimated from experimental Debye temperatures and DFT phonon calculations (neglecting minor dependencies of phonon frequencies for these solids on the choice of a solid state-appropriate GGA) were subtracted from the experimental cohesive energies as total solid atomization energy training targets.\cite{MCML}

\subsection{Atomic structures}

The above-described thermochemistry datasets QM9, W4-11, and W4-17 also provide optimized equilibrium geometries of the considered molecules and radicals. Staroverov {\it et al.}\cite{Staroverov2003} employed Gaussian-3X theory\cite{G3X} to determine the equilibrium bond lengths of 86 neutral molecules and 10 molecular cations (forming the T-96R dataset), and the equilibrium structures and vibrational frequencies for 69 neutral and 23 cation dimers (forming the T-82F dataset). These data can be used to train models to yield accurate equilibrium molecular geometries and harmonic vibrational properties.

For solids, the currently used benchmark and training data for structures are, as is the case for thermochemistry, based on experimental data with zero-point contributions removed from lattice constants and bulk moduli according to GGA phonon calculations or experimental Debye temperatures. Alchagirov {\it et al.}\cite{Alchagirov2001} used the latter approach entirely based on experimental data in a Debye model to remove zero-point contributions from the observed lattice constants and bulk moduli of 17 solids. Hao {\it et al.}\cite{Hao2012} computed the phonon frequencies of 58 cubic solids using the PBE GGA\cite{PBE} to remove zero-point contributions from the experimental lattice constants. Trepte {\it et al.}\cite{Trepte2022} similarly used PBE phonon calculations by Zhang {\it et al.}\cite{ZhangNJP2018} to subtract zero-point contributions from the experimental bulk moduli of 62 solids as training target for ML DFA models. The advantage of removing zero-point contributions from the experimental lattice and elastic data is that the DFA training can directly minimize the error on these properties with few total energy predictions without  also having to predict the computationally more costly phonon frequencies during training.

\subsection{Transition-metal surface chemistry}

Transition-metal surface chemistry constitutes an interesting challenge for DFAs, as spurious charge transfer problems between surface and adsorbates can occur and molecules can exhibit strong static correlation, while the metallic surface bandstructure requires a good description of metallic screening. Experimental results for 39 chemi- and physisorption energies on transition-metal surfaces were collected by Wellendorff {\it et al.}\cite{Wellendorff2015} and PBE zero-point contributions to be removed from the experimental results for DFA training were computed. These data were extended by one chemi- and one physisorption energy resulting in the ADS41 dataset.\cite{ADS41}

Surface reaction barrier heights pose the additional challenge of delocalization errors in transition state geometries for DFAs. Mallikarjun Sharada {\it et al.}\cite{sbh10} collected ten measured surface barrier heights for molecular dissociation on transition-metal surfaces forming the SBH10 dataset. This dataset was extended by Tchakoua {\it et al.}\cite{sbh17} by 7 more surface reaction barrier heights resulting in the SBH17 dataset.

These surface chemistry data are computationally more expensive to train ML DFA models against than the bulk solid data above. The considered bulk systems typically are cubic and highly symmetric, such that the lattice constant is the only structural degree of freedom to be optimized to predict equilibrium structures. For surface and adsorbate systems, nuclear coordinate degrees of freedom for surface and adsorbate atoms need to be relaxed, too, and the slab models typically contain on the order of 20 transition-metal atoms, thus rendering each total energy calculation significantly more expensive than in the small primitive unit cell bulk cases. These surface chemistry benchmarks from single-crystal experiments are, however, very valuable as their simulation requires accurate description of both extended and localized states and their interactions.

\subsection{Charge densities}

Medvedev {\it et al.}\cite{Medvedev2017} have benchmarked DFAs for their prediction of molecular charge densities compared to accurate quantum chemistry densities, finding that while newly developed DFAs have become better at predicting energies, the prediction of densities has been sacrificed to some degree. Quantum chemistry charge densities are thus an important training target for accurate ML DFAs. ML efforts reviewed in the following that consider such density metrics during their training typically computed quantum chemistry densities as part of the studies, with public quantum chemistry charge density benchmark data (rather than DFT densities) being hard to find.

\section{\sffamily \large ML XC FUNCTIONALS}

We divide ML DFAs into two categories. The first category summarized in the following consists of density functionals represented by mathematical expressions of explicit, tractable functional form with a low to moderate number of fitted numerical coefficients. The second category consists of functionals represented by neural networks with linear, convolutional, and non-linear activation layers,\cite{LeCun2015} where the neural network weights typically constitute a much larger space of fitting degrees of freedom than the coefficients in the former explicit density functionals forms.

\subsection{Semi-empirical DFAs with explicit functional forms}

A major (and lasting) impact on the adoption of GGAs for computational chemistry had the development of semi-empirical DFAs fitted against thermochemistry by Becke (followed a few years later by the non-empirical PBE\cite{PBE} functional and its lasting impact on computational materials science). In Becke's hybrid DFA B3PW91,\cite{b3pw91} relative weights for exact, local, semi-local exchange and local and semi-local correlation were fitted against the G2 thermochemistry dataset\cite{CurtissG2_1991} (the popular B3LYP functional replaces PW91\cite{PW91gradcorr} correlation with LYP\cite{LYP} correlation\cite{Stephens1994}; Vargas-Hernández\cite{vargashernandez2020} employed Bayesian optimization\cite{mockus_bayesian_1989} to choose the weighted exchange and correlation functionals and to optimize the weights). In Becke's B97 hybrid,\cite{B97} exchange and correlation inhomogeneity correction factors were introduced. These factors are polynomials in fractions of the reduced density gradient $\sim\!|{\bm \nabla}\rho|/\rho^{4/3}$, and the polynomial coefficients were also fitted against the G2 dataset. Mardirossian and Head-Gordon\cite{B97x-V} extended this work to a long-range screened hybrid and nonlocal correlation, $\omega$B97X-V, identifying which powers led to the best fit to a wider range of molecular benchmark data, also optimizing the empirical coefficients of the used VV10 nonlocal correlation functional,\cite{Vydrov2010} and eliminating a few fitting degrees with constraints. Liu {\it et al.}\cite{Liu2022} fitted a meta-hybrid with nonlocal correlation functional on a wide range of molecular chemistry benchmarks including thermochemistry, barrier heights, isomerization energies, excitation energies, non-covalent interactions, dipole moments, and bondlengths. 

Following the strategy of fitting polynomial coefficients to molecular thermochemistry, Hamprecht {\it et al.}\cite{Hamprecht1998} optimized inhomogeneity correction factors for a GGA with a small increase in number of polynomial coefficients compared to the Becke functionals, starting the development of the HCTH-family of XC functionals.\cite{HCTH1,HCTH2,HCTH3} Truhlar and co-workers developed the semi-empirical Minnesota XC functionals, which includes meta-GGAs and hybrid meta-GGAs.\cite{Minn1,Minn2,Minn3,Minn4,Minn5,Minn6} During the development of this XC functional family, the range of training data was widened, including solid properties in addition to molecular chemistry. In some of the Minnesota functionals, analytical constraints were used to reduce the large number (up to over 50) of fitting degrees of freedom.\cite{MinnesotaConstraints}

The performance of the strongly constrained and appropriately normed (SCAN) meta-GGA\cite{SCAN} clearly showed the benefits of fulfilling an increasing number of known analytical constraints for XC functionals.\cite{gedankendensities,Kaplan2023} Sparrow {\it et al.}\cite{Sparrow2022} expressed inhomogeneity factors for GGA exchange and correlation in a spline basis facilitating straightforward enforcement of equality and inequality constraints, resulting in the CASE21 functional for molecular chemistry. In addition to coefficient elimination for equality constraints, inequality constraints were implemented as penalties during exchange enhancement factor optimization of the meta-GGA MCML by Brown {\it et al.}\cite{MCML} for bulk, surface, and gas-phase chemistry. This constrained meta-GGA optimization was extended by Trepte {\it et al.}\cite{Trepte2022} to a simultaneously optimized, nonlocal VV10\cite{Vydrov2010} correlation term.

Rather than imposing analytical constraints, overfitting in the Bayesian error estimation functionals (BEEF) was suppressed by a quadratic (Tikhonov\cite{Tikhonov}) regularizer.\cite{mbeefvdw,mbeef,beefvdw} This ridge regression approach\cite{Hoerl1970} led to a fast decay of the magnitude of polynomial coefficients with increasing polynomial powers in the series used to expand the GGA and meta-GGA exchange enhancement factors. In contrast to many non- and semi-empirical XC functionals, the BEEF functionals enhance exchange for a homogeneous system over the exact local exchange in this limit. Similar exchange enhancement increase in the homogeneous limit by a few percent was also observed by Kovács {\it et al.}\cite{Kovacs2022} for meta-GGA exchange enhancement fits to lattice constants, solid cohesive energies, and band gaps when not imposing this LDA limit.

Generally, the XC functional fitting approaches have evolved into using generic forms of enhancement or inhomogeneity factors with a large number of fitting degrees of freedom. Gastegger {\it et al.}\cite{Gastegger2019} instead applied genetic programming\cite{Koza1994} to find mathematically simple forms of XC functionals performing well for molecular chemistry benchmarks. Ma {\it et al.}\cite{Ma2022} used similar symbolic regression techniques (by means of regularized evolution\cite{Real2019}) to evolve generations of XC functionals starting from preexisting ones to new ones with improved performance on target datasets.  

The above-reviewed semi-empirical XC functionals are optimized to yield quantitatively improved performance for desired chemistry target metrics. Fundamental DFA issues of the different levels of theory of hybrids, GGAs, and meta-GGAs are generally unlikely improved upon qualitatively by numerical optimization of these mathematically relatively simple XC functionals. Much more complex DFAs, such as neural network-based approaches reviewed below, might offer an opportunity to arrive at such qualitative improvement. Accurate, but system or materials class-specific functionals could also be learned with simpler ML models. Riemelmoser {\it et al.}\cite{Riemelmoser2023} used Gaussian kernel regression with nonlocal density features to learn ML functionals reproducing RPA correlation energies for diamond and water, respectively.

\subsection{Neural network DFAs}

\begin{figure}
\centering
\includegraphics[width=3.33in,keepaspectratio=true]{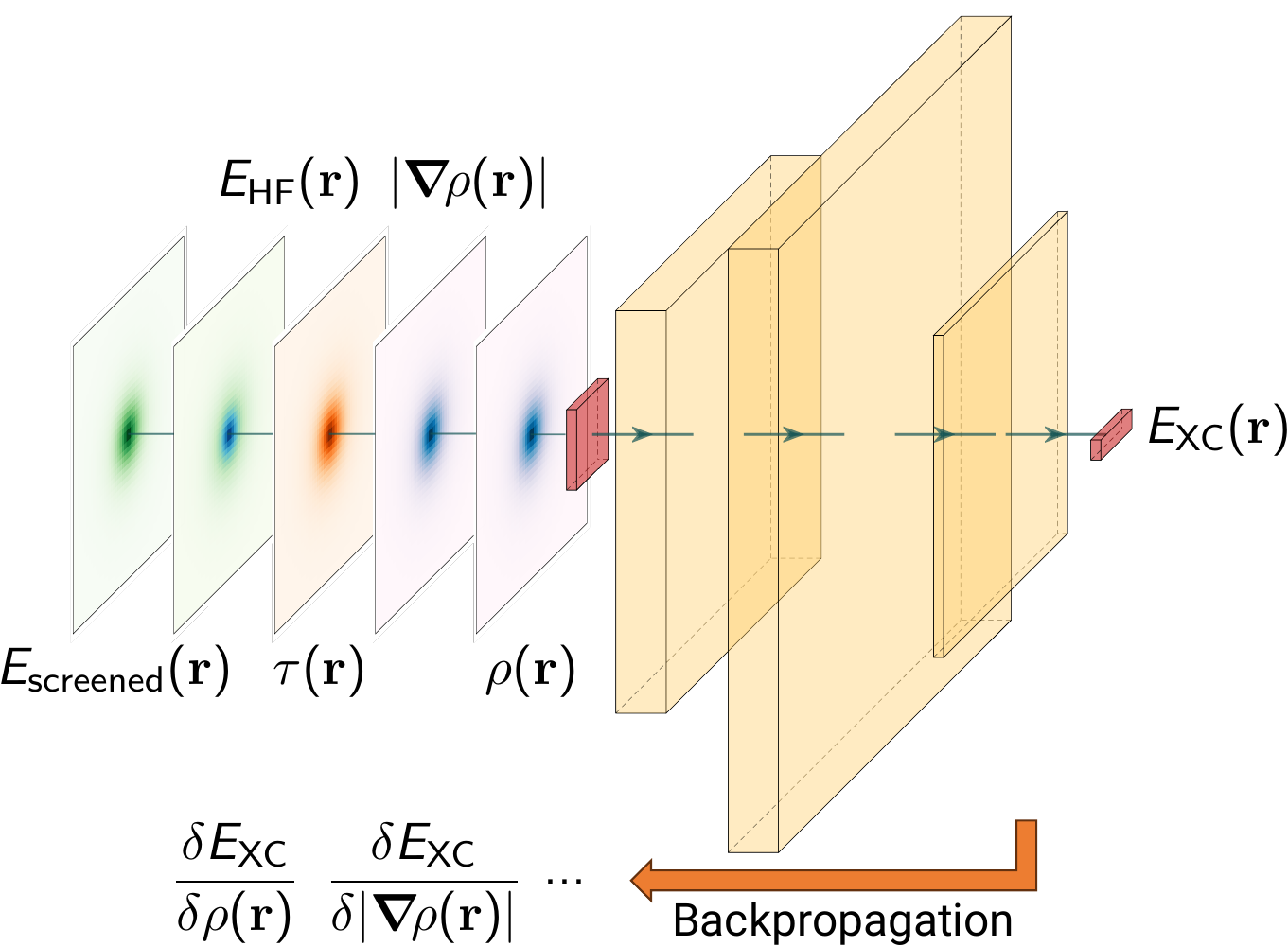}
\caption{\label{fig:neuralxc} Schematic of a neural network-based XC functional. Local features of the charge density $\rho$ at position {\bf r} and, depending on the XC functional type, kinetic ($\tau$) or EXX energy densities are inputs to the neural network yielding the XC energy $E_{\rm XC}(\textbf{r})$. With the help of backpropagation, the gradient of the XC energy with respect to the inputs can be obtained, from which the effective one-body KS potential is computed.}
\end{figure}

Neural network XC functionals are functionals of density, density gradient and additional local electronic properties, such as KS kinetic and EXX energy densities, and are trained to yield accurate target XC energies and typically also accurate ground state densities (Figure~\ref{fig:neuralxc}). The automatic differentiation-based backpropagation technique\cite{backprop} commonly used for optimizing the neural network weights for loss function minimization during training is here applied to also compute the gradient of the XC energy with respect to the local density and energy density inputs. These derivatives are required to compute the KS potential.

\begin{figure}
\includegraphics[width=\textwidth,keepaspectratio=true]{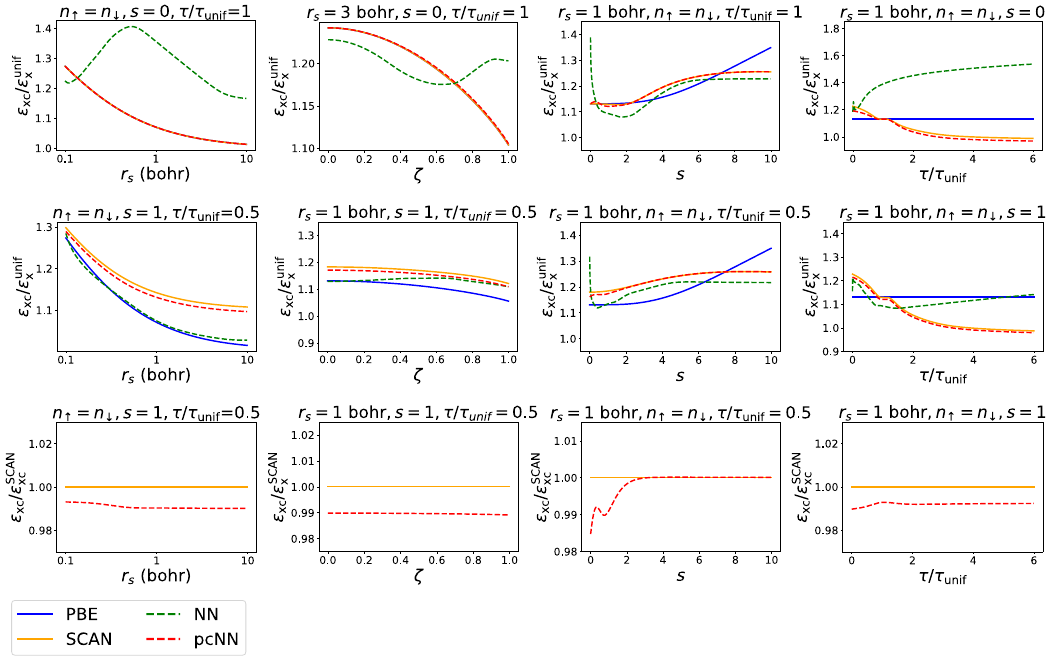}
\caption{\label{fig:pcNN} Comparison of the XC enhancement of the PBE,\cite{PBE} SCAN,\cite{SCAN} and neural network-based XC functionals without (NN) and with physical constraints (pcNN) for different values of Wigner-Seitz radius $r_s$, reduced density gradient $s$, KS kinetic energy density $\tau$ (relative to Thomas-Fermi kinetic energy density $\tau_{\rm unif}$), and relative spin polarization $\zeta$. Two top rows: XC enhancement over local exchange $\epsilon_{\rm X}^{\rm unif}$, bottom row: pcNN XC enhancement over SCAN XC functional. Reproduced from Ref.~\citeonlb{pcNN}%
, \protect\urllink[border = 0, color = 0 0 1]{https://doi.org/10.1103/PhysRevResearch.4.013106}{{\color{blue}DOI: 10.1103/PhysRevResearch.4.013106}}
\ under the terms of the Creative Commons Attribution 4.0 International License. Copyright 2022, the Authors. Published by the American Physical Society.}
\end{figure}

Nagai {\it et al.}\cite{Nagai2020} trained a neural network meta-GGA on the quantum chemistry densities and atomization energies of only the three molecules H$_2$O, NH$_3$, and NO. The neural network weights were stochastically optimized by selfconsistently computing the ground state densities and energies of the three molecules and H, N, and O atoms for randomly perturbed weights. Remarkably, the resulting XC functional performs well and even outperforms existing functionals across benchmark databases containing over 100 molecules. The implied transferability from a very small training set to other molecules and also to reaction barriers is very promising. This neural network meta-GGA work was extended by enforcing five physical constraints on exchange and correlation each, and including the ionization potentials as an additional training target besides densities and atomization energies of the molecules H$_2$O, NH$_3$, and CH$_2$.\cite{pcNN} This XC functional was trained as an XC enhancement over the SCAN meta-GGA. While the resulting XC functional bares some similarity with the SCAN functional (Figure~\ref{fig:pcNN}), it is reported to outperform SCAN on lattice constant and molecular atomization benchmarks.

Kirkpatrick {\it et al.}\cite{Kirkpatrick2021} developed a neural network XC functional addressing the total energy {\it vs.}\ fractional particle number and derivative discontinuity problem. Their DeepMind21 (DM21) functionals take point by point in real space charge density and gradient, KS kinetic, unscreened and long-range EXX energy densities as inputs. The training was performed non-selfconsistently on B3LYP densities and KS orbitals. The loss was computed as the energy differences to large sets of molecular chemistry benchmarks. The change in energy a single selfconsistent field iteration would cause starting from the B3LYP orbitals was estimated perturbatively, and this change was penalized as an additional term in the loss function. With this approach, the neural network weights could be optimized using the gradient obtained via backpropagation, as no KS selfconsistency cycles were required during training. This enabled training a neural network with $\sim\!4\cdot10^5$ trainable weights (Nagai {\it et al.}\cite{Nagai2020} used $\sim\!2\cdot10^4$ trainable weights) on large datasets, which were extended by densities of fractional charges and spins to train the correct piece-wise linear behavior with energy {\it vs.}\ particle number derivative discontinuities. Energies and charge densities were interpolated linearly between integer particle numbers as dictated by quantum mechanical ensemble averages, spin densities were interpolated linearly between degenerate spin states, and the KS inversion technique due to Wu and Yang\cite{Wu2003} was used to decompose the interpolated densities into KS orbitals for computation of the required DM21 input energy densities. Several variants of the DM21 functional were developed with different imposed constraints (fractional charge and spin constraints and homogeneous electron gas limit), showing promising performance on molecular benchmark data. Tests of two of these variants for solids, {\it i.e.}, materials classes outside the training data, are presented in the outlook at the end of this review.

Gedeon {\it et al.}\cite{Gedeon2022} trained a neural network XC functional for correct derivative discontinuity behavior for 1D systems with the help of an additional neural network input feature explicitly depending on the total (fractional) particle number. Wang {\it et al.}\cite{Wang2023} iteratively trained a neural network GGA by using backpropagation to optimize the neural network at fixed charge densities and then using this neural network GGA to recompute charge densities in KS selfconsistency cycles at fixed neural network weights. Chen {\it et al.}\cite{ChenDeePKS2021} similarly performed iterative loops of neural network optimization at fixed orbitals and KS solution at fixed neural network weights for DFT and hybrid DFT. 

Rather than employing the techniques in the above approaches to XC neural network optimization using stochastic approaches, perturbative approaches, and iterative loops between neural network backpropagation-based optimization and KS selfconsistency cycles, there are efforts to make the KS solution with its challenge of required selfconsistency between KS orbitals and density differentiable. Li {\it et al.}\cite{LiPRL2021} implemented a differentiable KS solution for 1D systems, which they trained on accurate solutions possible in this reduced dimensionality within the density matrix renormalization group (DMRG).\cite{DMRG} Their loss function penalizes differences in the converged KS density and the DMRG one as well as the total energy difference at each KS iteration to the DMRG energy. This approach enhances the smooth convergence rate of the KS cycles towards the DMRG solution, with the KS equations effectively acting as a regularizer. The KS regularization helps the ML functional learn smooth functional derivatives not only at the converged solution, but along the KS convergence trajectory. The regularization was furthermore found to enhance the transferability of the XC functional.\cite{Kalita2022} Kasim and Vinko\cite{KasimPRL2021} implemented differentiable KS solutions in three dimensions, and Kasim {\it et al.}\cite{KasimDGC2022} extended this work to a differentiable DFT and HF code. Starting from accurate ground state densities instead, Tozer {\it et al.}\cite{Tozer1996} used a neural network to model the XC potential derived from configuration-interaction densities of atoms and small molecules using the KS inversion technique by Zhao {\it et al.}\cite{Zhao1994} for finite systems. 

\section*{5\ \ \ \sffamily \large $\boldsymbol\Delta$-ML CORRECTIONS TO DFT}

\addcontentsline{toc}{section}{\sffamily \large Delta-ML CORRECTIONS TO DFT}

\setcounter{section}{5}

The approaches reviewed in the previous section learn approximations to the XC functional $E_{\rm XC}[\rho]$, and functional differentiation yields the effective one-body KS potential used to compute KS orbitals and thus KS kinetic energies, EXX integrals, and other orbital-dependent energies. Instead of using the KS potential from the machine-learned XC functional and correspondingly solving the KS equations selfconsistently, the $\Delta$-ML methods reviewed here provide post-DFT and post-hybrid DFT corrections to $E_{\rm XC}[\rho]$, where the ground state density $\rho$ and KS orbitals are kept fixed at the solution from another XC functional.

A resulting simplification for such $\Delta$-ML methods is that they need not provide functional derivatives, as one-body KS potentials are not computed. This enables the usage of non-differentiable ML approaches such as the decision tree ensemble method of gradient boosting.\cite{Breiman1999,Friedman2001} Wang {\it et al.}\cite{WangXGBoost2022} used the popular gradient boosting implementation XGBoost\cite{XGBoost} to learn a density and density gradient-dependent ({\it i.e.}\ GGA-type) XC functional to be evaluated non-selfconsistently on PBE charge densities. This functional provides an XC energy correction on top of PBE XC, which was trained to improve upon molecular thermochemistry benchmarks. Similarly, Bogojeski {\it et al.}\cite{Bogojeski2020} used kernel ridge regression to train a model correcting selfconsistent PBE-based energies to coupled cluster results. This XC model takes nonlocal density features as input (a second kernel ridge regression model was developed to predict these density features from the atomic structure alone). They found these $\Delta$-XC energy corrections could be regressed with lower errors than coupled cluster or DFT energies directly, constituting another advantage of $\Delta$-ML corrections to XC energies.

Another family of approaches non-selfconsistently provides correlation energies post selfconsistent DFT or HF. Margraf and Reuter\cite{Margraf2021} developed a kernel-based approach using nonlocal charge density features that provides non-selfconsistent correlation energies trained on quantum chemistry energies. Chen {\it et al.}\cite{ChenDeePHF2020} developed a neural network-based approach for post-HF correlation energies also trained on molecular quantum chemistry. They used the local one-body density matrix as model input and trained an ensemble of neural networks. Deviation in the ensemble was used as a model uncertainty estimate in an active learning approach to limit the number of computationally expensive quantum chemistry simulations for training data. Cheng {\it et al.}\cite{Cheng2019} used Gaussian process regression\cite{GPR2006} to develop a post-HF model for correlation energies using the one-body density matrix. They learned a density matrix functional of features of the HF orbitals and found good transferability of the model from training on the QM7 dataset\cite{qm7} to significantly larger molecules from the GDB-13\cite{gdb13} dataset. One-body density matrix-derived features were similarly used by Ng {\it et al.}\cite{Ng2023} to train neural networks to predict post-HF correlation energies, finding good transferability from training on small water clusters to predictions on large water clusters and from training on short alkanes to predictions on longer alkanes. Han {\it et al.}\cite{HanMLMP2021} devised a semi-local ML XC functional for post-HF correlation energies trained against M{\o}ller-Plesset-2\cite{MoellerPlesset} data. Their functional takes charge density and gradient, electronic kinetic energy, and a weighted sum of the occupied orbital densities as local input. For the latter sum, the orbitals were weighted reciprocally with the energy difference of the HF energy level of the occupied orbital to the virtual orbitals, in order to mimic the energy denominators in second order perturbation theory in this feature.

\section{\sffamily \large ATOMIC STRUCTURE-DEPENDENT XC CORRECTIONS}

\begin{figure}
\centering
\includegraphics[width=3.33in,keepaspectratio=true]{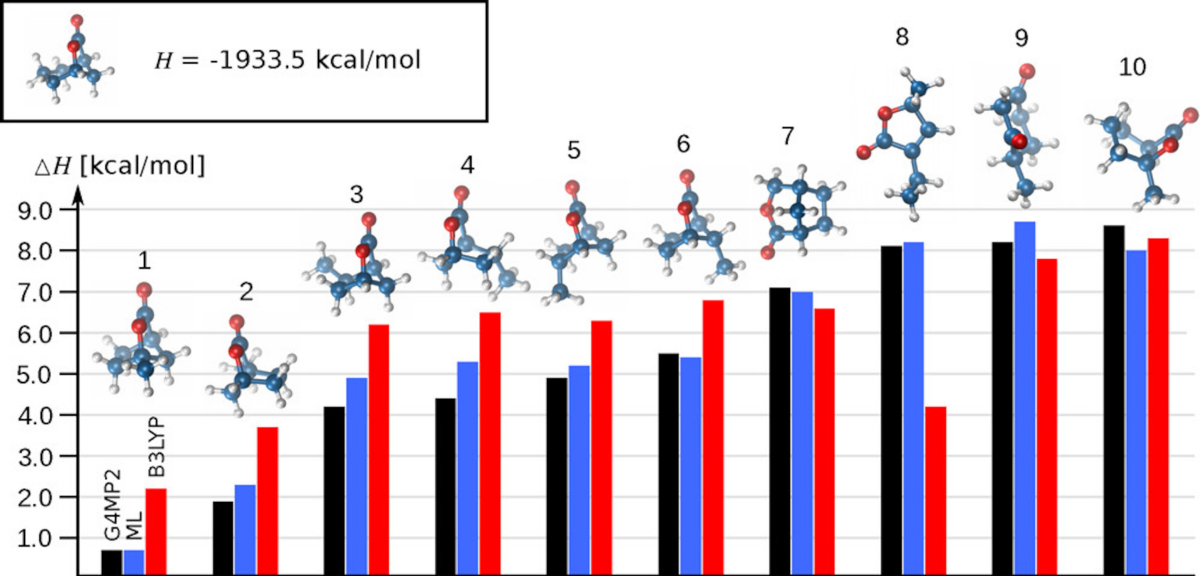}
\caption{\label{fig:Enthalpies} Reaction enthalpies of lowest energetic C$_7$H$_{10}$O$_2$ isomers with respect to the most stable isomer 7-oxabicyclooctan-7-one (depicted in inset). Gaussian-4-M{\o}ller-Plesset-2 benchmarks are shown by black bars and B3LYP DFT results by red bars. Blue bars show $\Delta$-ML predictions based on atomic-structural kernels significantly improving over the B3LYP results to within chemical accuracy $<1$~kcal/mol. Reproduced with permission from Ramakrishnan {\it et al.},\cite{Ramakrishnan2015}%
\ \protect\urllink[border = 0, color = 0 0 1]{https://doi.org/10.1021/acs.jctc.5b00099}{{\color{blue}DOI: 10.1021/acs.jctc.5b00099}}.
 Copyright 2015 American Chemical Society.}
\end{figure}

While the $\Delta$-ML methods summarized in the previous section predict post-DFT and HF corrections based on electronic features, atomic structural information can be employed to parameterize such XC corrections, too. Ramakrishnan {\it et al.}\cite{Ramakrishnan2015} used kernel ridge regression to non-selfconsistently correct HF to M{\o}ller-Plesset-2 and coupled cluster results as well as M{\o}ller-Plesset-2 to coupled cluster. The corrections are expressed as sums over terms measuring atom by atom the similarity of two structures through kernel functions exponentially decaying with atomic coordinate distances. A resulting ML-model was shown to improve B3LYP-based reaction energies of organic molecules to chemical accuracy, both correcting over- and severely underestimated DFT reaction energies (see Figure~\ref{fig:Enthalpies}).

The DFT+$U$ method constitutes an atomic structure-dependent XC correction, as the Hubbard terms in this method are used to penalize fractional occupation of local density matrices in the basis of, {\it e.g.}, correlated transition-metal or rare-earth site-centered atomic orbitals. Using genetic programming with experimental transition-metal oxide heats of formation as training data, $\Delta$-ML models were found that featurizing local density matrices enable reaction energy predictions based on different levels of theory for products and reactants.\cite{Voss2022} Localized states in the correlated oxides were treated at the DFT+$U$ level with site-dependent, first-principles $U$-parameters from linear-response theory, and delocalized states in the metallic phases were treated at the DFT level.

Empirical force fields are ubiquitously used to supplement semi-local DFAs with dispersion energetics. Proppe {\it et al.}\cite{Proppe2019} used Gaussian process regression featurizing the individual pairwise terms of such vdW force fields to arrive at improved agreement with coupled cluster results. Variance prediction was used to select systems for coupled cluster benchmark calculations in an active learning framework.

\sloppy

A non-selfconsistent $\Delta$-ML approach with an atomic structure and charge density-dependent XC energy correction was developed by Dick and Fernandez-Serra.\cite{Dick2019} In this work, the selfconsistent DFT charge density was projected onto atom-centered basis functions, defined as spherical harmonics times a radial function confined to a spherical shell around the atom. In analogy to ML inter-atomic potentials, the $\Delta$-XC energy correction was defined as a sum over atomic contributions, where each of these contributions is modeled by a neural network with the charge projections around the corresponding atom as input. As this non-selfconsistent $\Delta$-ML approach does not provide corrections to the charge density, corrections to the atomic forces were trained with additional neural networks. The models were trained and tested on water clusters, for which in turn coupled cluster-based force field parameterizations\cite{Babin2013,Babin2014} were used as reference data.

\fussy

\begin{figure}
\centering
\includegraphics[width=3.33in,keepaspectratio=true]{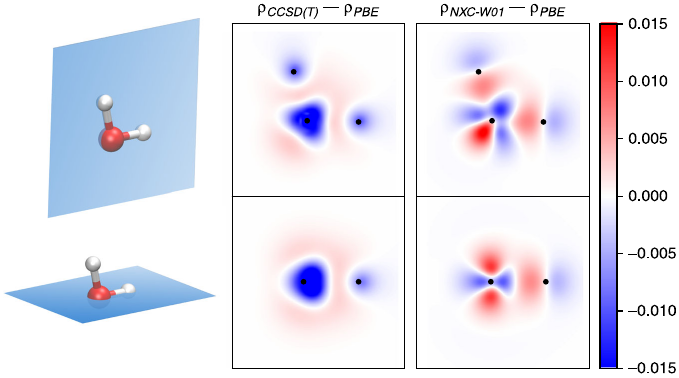}
\caption{\label{fig:Water} Comparison of water molecule charge density differences between coupled cluster theory and PBE (contour plots to the left) and between the NeuralXC functional and PBE (right). Reproduced from Ref.~\citeonlb{Dick2020}%
, \protect\urllink[border = 0, color = 0 0 1]{https://doi.org/10.1038/s41467-020-17265-7}{{\color{blue}DOI: 10.1038/s41467-020-17265-7}}%
\ under the terms of the Creative Commons Attribution 4.0 International License. Copyright 2020, the Authors. Published by Springer Nature.}
\end{figure}

Dick and Fernandez-Serra\cite{Dick2020} extended this work to refined atom-projected density features that account for the charge density difference of the system to atomic reference densities, and they used backpropagation to compute derivatives with respect to these density features. They thus compute the functional derivative of their model with respect to the density resulting in an ML XC functional that can be used selfconsistently. Again benchmarking their model for water, they find, {\it e.g.}, an improved description of charge accumulation along the OH bonds compared to semi-local DFAs (Figure~\ref{fig:Water}).

\section{\sffamily \large ML KS HAMILTONIAN SUBSTITUTIONS}

ML approaches are not only used to find approximations to $E_{\rm XC}[\rho]$, they can also be employed, {\it e.g.}, to approximate the non-interacting kinetic energy functional $T_{\rm KS}[\rho]$ in Eq.~\ref{eq:kohnsham}. Such methods are not necessarily aimed at increasing the accuracy of DFT simulations, but typically rather at their acceleration. An explicit density-only functional for the non-interacting kinetic energy in KS orbital-free DFT would substantially reduce the computational cost of DFT, which for semi-local DFT with $T_{\rm KS}$ computed from KS orbitals is dominated by the diagonalization of the auxiliary Hamiltonian with KS one-body potential. Burke and coworkers developed neural network models for $T_{\rm KS}[\rho]$ for 1D systems.\cite{Snyder2012,LiIJQM2016} Tan {\it et al.}\cite{Tan2023} implemented a differentiable, orbital-free DFT code for efficient training of neural network-based approximations to $T_{\rm KS}[\rho]$.

\begin{figure}
\centering
\includegraphics[width=3.33in,keepaspectratio=true]{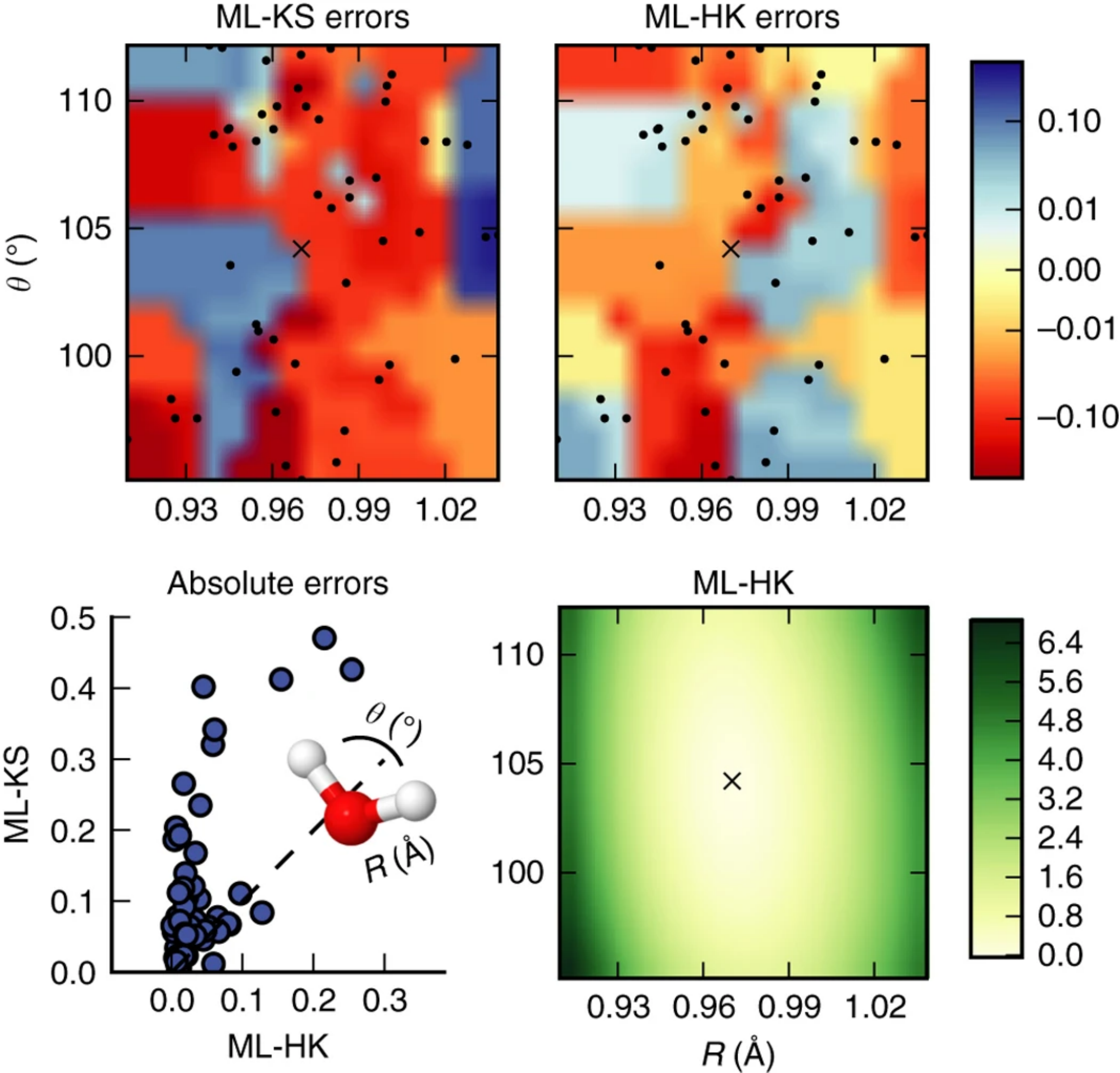}
\caption{\label{fig:MLHK} Performance of ML external potential to energy (ML-KS) and external potential to electronic charge density mappings (ML-HK) for H$_2$O (all energies in kcal/mol). Top: deviation from PBE-DFT energies as a function of averaged bond length and angles, left for ML-KS, right for ML-HK. Bottom left: errors with respect to PBE-DFT energies for test set geometries. Bottom right: ML-HK potential energy surface of H$_2$O with minimum in agreement with PBE-DFT (marked by cross). Reproduced from Ref.~\citeonlb{Brockherde2017}%
, \protect\urllink[border = 0, color = 0 0 1]{https://doi.org/10.1038/s41467-017-00839-3}{{\color{blue}DOI: 10.1038/s41467-017-00839-3}}%
\ under the terms of the Creative Commons Attribution 4.0 International License. Copyright 2017, the Authors. Published by Springer Nature.}
\end{figure}

Finding that functional derivatives of neural network models of $T_{\rm KS}[\rho]$ were typically noisy and not generally useful for implementing a minimization scheme of the KS orbital-free energy functional (and regularization of the functional derivatives unfortunately correlated with a loss of accuracy of $T_{\rm KS}[\rho]$\cite{LiIJQM2016}), Brockherde {\it et al.}\cite{Brockherde2017} machine-learned the map from external potential to ground state charge density  instead of learning to approximate $T_{\rm KS}[\rho]$. This approach absolves from a total energy functional minimization with respect to the density, and functional derivatives are thus not required. It was furthermore found that the approach of predicting the charge density yielded lower errors with moderate amounts of training data than learning to directly predict the energy of the system from the external potential (see Figure~\ref{fig:MLHK}). Given the promise of significantly improved computational efficiency of such an orbital-free DFT approach, further ML models directly predicting the charge density have been developed.\cite{Grisafi2019,Fabrizio2019,Chandrasekaran2019,ZepedaNunez2021,DelRio2023,Focassio2023} Shao {\it et al.}\cite{Shao2023} furthermore learned maps from external potential to the one-body density matrix.

Another avenue for electronic structure ML models targeting computational efficiency is the substitution of expensive energy contributions, such as EXX, with cheaper ML approximations. Cuierrier {\it et al.}\cite{Cuierrier2022} used a neural network to train a semi-local approximation of exchange exploiting the fourth order expansion of the exchange hole\cite{PhysRevA.96.022502} as an input feature, where a relatively small neural network was optimized with a quasi-Newton approach. Their training target was EXX from hybrid DFT. Lei and Medford\cite{LeiPRM2019} trained a neural network to reproduce B3LYP XC with nonlocal density descriptors. Yu {\it et al.}\cite{YuNCM2020} used Bayesian optimization to find Hubbard-$U$ corrections that best reproduce the behavior of EXX in short-range screened hybrid DFT for solids. Bystrom and Kozinski\cite{Bystrom2022} used Gaussian process regression to train an exchange functional reproducing EXX with a combination of semi-local and nonlocal density features. They find that their ML model could be used as a computationally cheaper replacement for EXX in hybrid functionals and show good agreement with hybrid DFT for thermochemistry and ioniziation potentials. Overall, these approaches aim at reproducing the effects of EXX on the electronic structure at the cost of semi-local DFT.

\section{\sffamily \large CHALLENGES AND OPPORTUNITIES}

The perhaps surprising usefulness of the simplest of DFAs, the LDA, even for inhomogeneous systems, was explained by XC energies only depending on the spherical average of the XC hole and the LDA fulfilling the important sum rule of this hole accounting for one missing electron.\cite{GunnarssonPRB1976} Density gradient corrections do not necessarily improve upon the predictive performance of the LDA, unless this sum rule is fulfilled. It was the real-space cut-off in the exchange hole gradient expansion introduced by Perdew\cite{PerdewPRL1986} that enabled the development of GGAs fulfilling the sum rule and outperforming the LDA for energetics and atomic structures. The success of the seemingly crude LDA and the reason for sum rule-fulfilling GGAs being able to outperform the LDA can thus be rationalized through physical understanding.

ML models, on the other hand, do not necessarily allow for such insights. Even with the simpler reviewed ML models with only a moderate number of fitting degrees of freedom, it cannot necessarily be excluded that the improvement over existing approaches merely consists of reducing error bars on the well-known benchmark data sets,\cite{Becke2022} potentially at the price of {\it e.g.}\ sacrificing the description of the central quantity of DFT, the charge density, for improved reaction energy differences,\cite{Medvedev2017} and thus potentially developing XC functionals that yield ``right answers'' for the ``wrong reason''.\cite{HammesSchiffer2017} These empirical methods could show accidentally improved energetics with respect to some benchmarks without having led to improved physical approximation of electronic XC.

\begin{figure}
\centering
\includegraphics[width=3.33in,keepaspectratio=true]{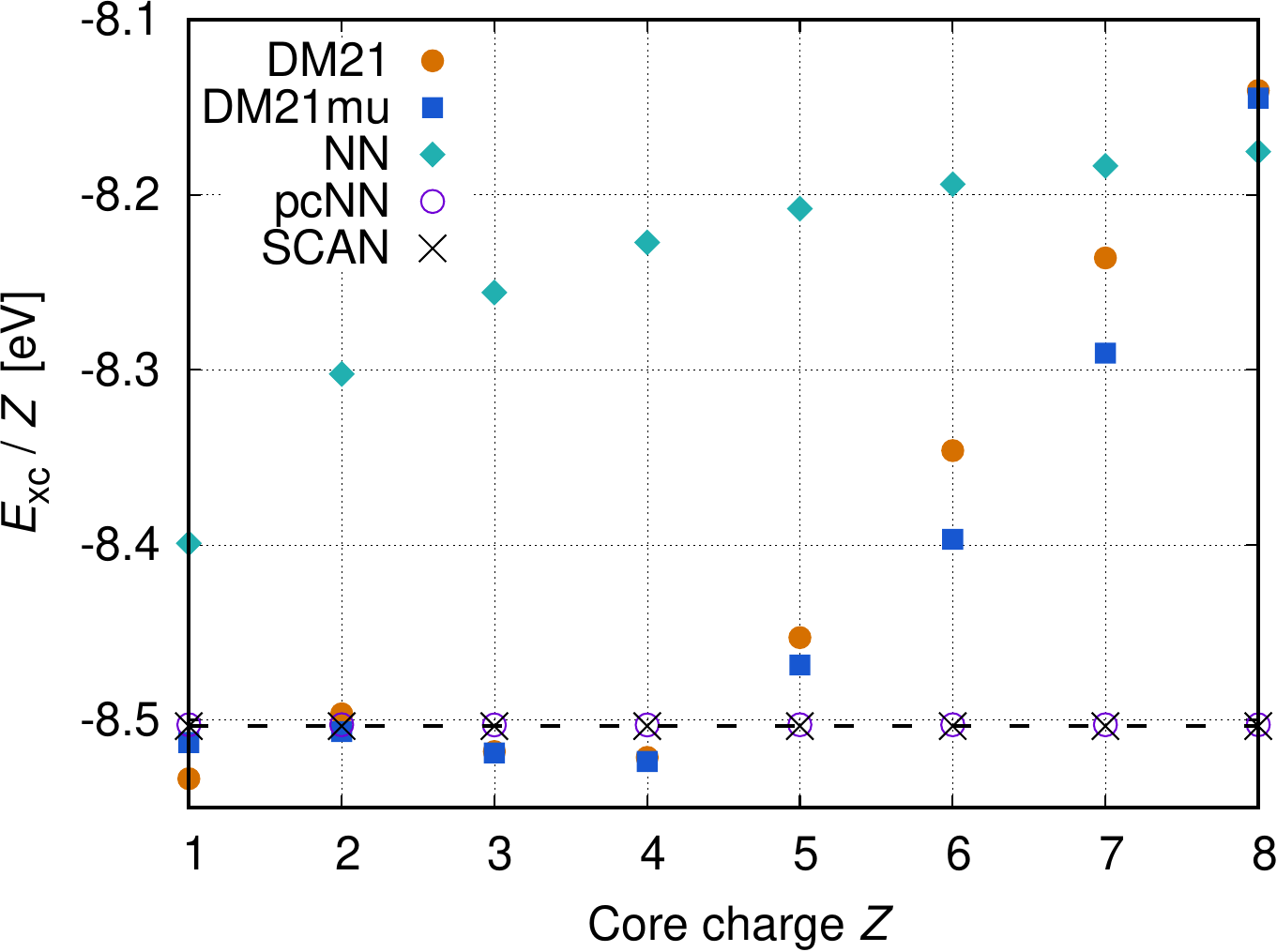}
\caption{\label{fig:Hydrogenic} XC energies evaluated on the analytical exponential densities of hydrogenic ions as a function of the core charge $Z$. The dashed line marks the exact exchange energy of \mbox{$-5/16$ Hartree $\cdot$ $Z$}, which exactly cancels the spurious Hartree interaction in these one-electron systems.}
\end{figure}

To highlight the promise that ML XC models hold, but also challenges in their transferability, we will compare here the performance of ML XC models to systems far outside their training and validation data. We begin this discussion with a simple test of the XC models for their performance on hydrogenic ions (Figure~\ref{fig:Hydrogenic}). In these one-electron systems, accurate XC should exactly cancel the spurious one-electron Hartree interaction. The first neural network XC model (NN) by Nagai {\it et al.},\cite{Nagai2020} only trained on three molecules but showing promising performance for a range of molecular thermochemistry, did not explicitly incorporate analytical constraints nor was it trained {\it e.g.} on the exact XC density of the H atom. Its performance for hydrogenic ions turns out be relatively poor. The physically constrained pcNN\cite{pcNN} by the same authors, also only trained on three molecules, yields accurate XC energies for these ions. From the DM21 family of functionals, we consider the variant DM21 with fully imposed fractional charge and spin piece-wise linear energy behavior and the variant DM21mu with imposed homogeneous electron gas limit. Both DM21 and DM21mu show reasonable accuracy for hydrogenic ions with $Z<5$, where the more positively charged ions are likely too far from the training data.

\begin{figure}
\centering
\includegraphics[width=3.33in,keepaspectratio=true]{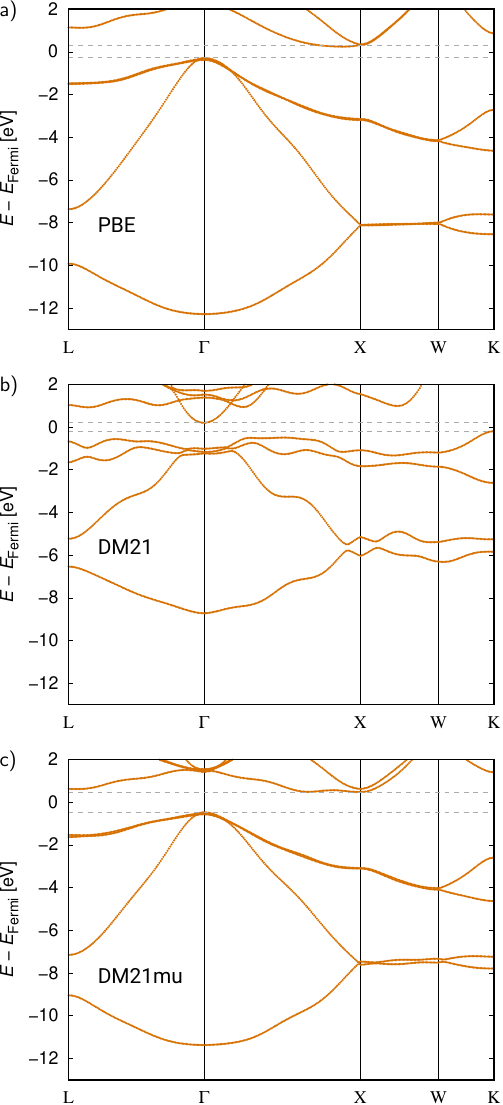}
\caption{\label{fig:SiBS} KS bandstructure of Si: a) computed with the PBE functional; b) and c) computed non-selfconsistently using PBE KS orbitals with the DM21 and DM21mu functionals, respectively. The dashed lines indicate the valence and conduction band edges. While PBE and DM21 significantly underestimate the Si band gap, DM21mu yields a good result of $\sim$1~eV.}
\end{figure}

The derivative discontinuity trained for in the DM21 functional via fractional total charges on molecular systems is of importance for correcting KS gaps to the physical, fundamental gap of semiconductors and insulators. We thus test here the performance of DM21 functionals trained only on molecular data for a bulk system: the semiconductor Si \mbox{(Figure~\ref{fig:SiBS}).} PBE, as is typical for DFAs, significantly underestimates the Si band gap. The DM21 functional shows poor performance for the Si bandstructure in general. The spurious oscillations in band dispersion as a function of wave vector are likely due to DM21 not being parameterized in the density and energy gradient neural network input tuples relevant for solids here. The bandstructure is overall compressed in energy range and with that the band gap, while one would have hoped that a functional reproducing derivative discontinuities would yield an enlarged band gap with respect to semi-local DFAs. DM21mu, on the other hand, yields a smooth band structure and an increased band gap in approximate agreement with experimental and $GW$\cite{Hedin1965} results.\cite{Hybertsen1986} This is another example of the importance of physical constraints on the ML XC models: although DM21mu was only trained on molecular systems, the homogeneous electron gas constraint seems to have extended the range of applicability over DM21 significantly.

\vfill

\begin{figure}[b!]
\centering
\includegraphics[width=3.85in,keepaspectratio=true]{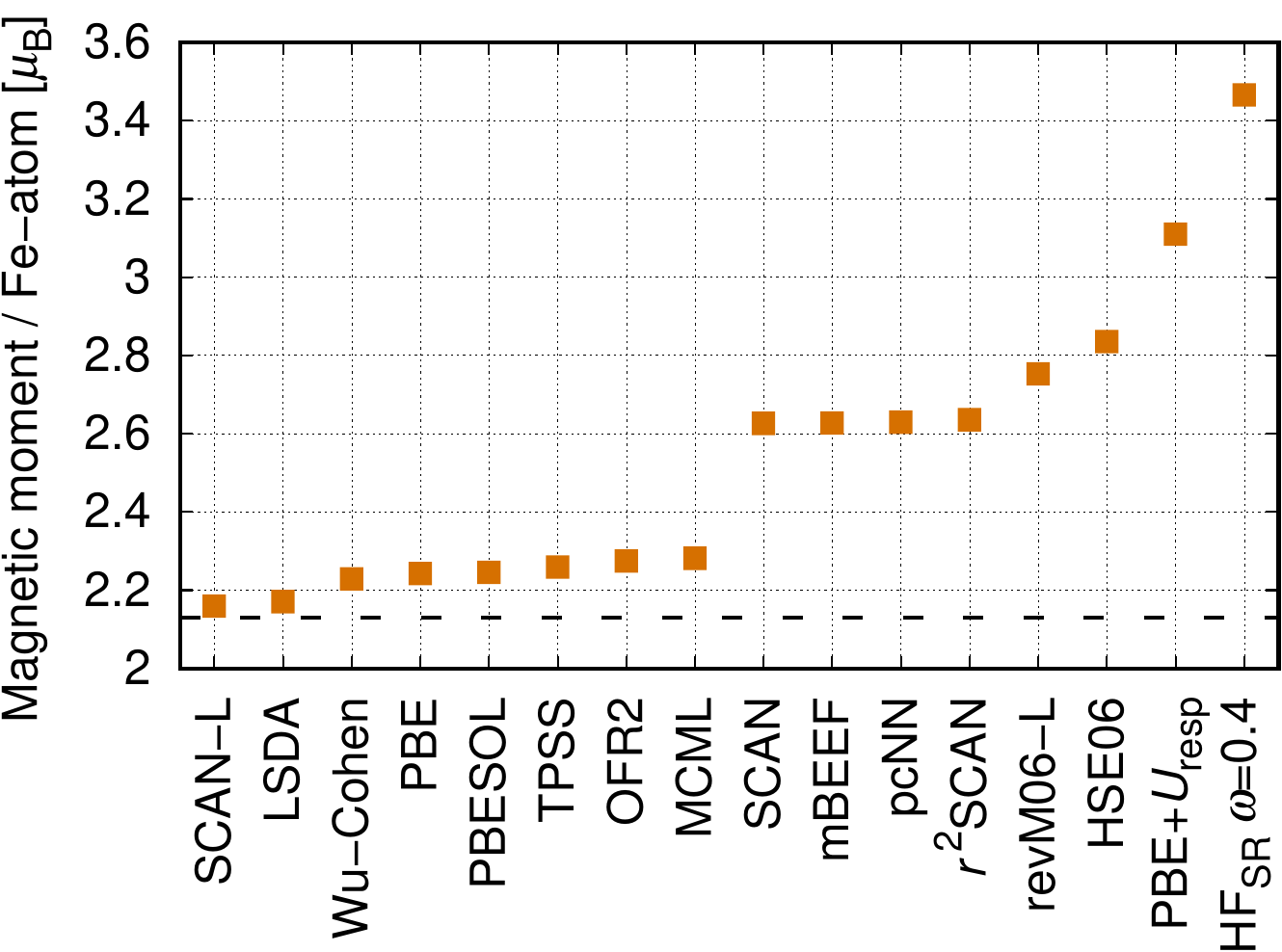}
\caption{\label{fig:Iron} Magnetic moment per primitive body-centered cubic unit cell of bulk Fe obtained with different DFAs: SCAN-L,\cite{SCANL} LSDA,\cite{PerdewZunger1981} Wu-Cohen,\cite{wucohen} PBE,\cite{PBE} PBEsol,\cite{pbesol} TPSS,\cite{TPSS} OFR2,\cite{OFR2} MCML,\cite{MCML} SCAN,\cite{SCAN} mBEEF,\cite{mbeef} pcNN,\cite{pcNN} $r^2$SCAN,\cite{R2SCAN} revM06-L,\cite{REVM06L} and HSE06.\cite{hse06} PBE+$U_{\rm resp}$ result with first-principles Hubbard parameter from linear response from \mbox{Ref.~\citeonl{Voss2022}}. HF$_{\rm SR}\,\omega$=$0.4$ is short-range screened exchange with the same inverse screening length of $0.4/{\rm Bohr}$ as for long-range screened exchange used as input feature in the DM21 functionals. The dashed line indicates an experimental value of $2.13$~$\mu_{\rm B}$.\cite{Wijn1997}}
\end{figure}

\pagebreak

It is interesting to note that DM21mu reduces the $sp$-bandwidth of Si compared to PBE. DFAs have a general problem of overestimating bandwidths.\cite{PhysRevLett.59.819} Using bandstructures determined from angle-resolved photoemission spectra as accurate training data for ML DFAs could be an option to incorporate the underlying many-body effects into the XC functional (if one chooses to interpret the KS eigenlevels as quasi-particle energies). If DM21mu is transferable to semiconductors in general, will need to be tested on more systems. Extending training data to $GW$ quasi-particle bandstructures could be of help to teach the functional derivative discontinuities explicitly for solids (and this would likely cure minor flaws in the DM21mu Si bandstructure such as the lowered conduction band at the L-point).

With these very encouraging results for DM21mu for a semiconductor bandstructure, we now turn to a problem where beyond semi-local DFT approaches typically perform worse than the LDA and GGAs: itinerant ferromagnetism. Figure~\ref{fig:Iron} shows the magnetic moment of bulk Fe computed with a number of different analytical, semi-empirical, and ML DFAs and beyond semi-local DFT methods. Unfortunately, the majority of advanced XC functionals shown here are not able to reproduce the metallic screening of exchange sufficiently and correspondingly yield too large magnetic moments. This problem was addressed in the strongly constrained SCAN functional by deorbitalization:\cite{PhysRevB.100.041113} the KS kinetic energy dependence was replaced by a density functional for the kinetic energy involving the Laplacian of the charge density, and the resulting SCAN-L\cite{SCANL} and OFR2\cite{OFR2} show significantly improved magnetic moments, albeit at the prize of sacrificing constraint fulfillment. Unfortunately, higher rungs of theory such as the short-range screened hybrid HSE06\cite{hse06} do not lead to an improved description here. Employing a Hubbard-$U$ term to compensate the spurious GGA curvature for fractional electron count even increases the magnetic moment further. Finally, we computed the Fe magnetic moment with short-range screened HF with the same screening length used for long-range screened EXX in the DM21 and DM21mu features, leading to largely exaggerated magnetic moments. Unfortunately, both DM21 and DM21mu show such exaggeratedly large magnetic moments, too (roughly estimated from differences in the spin densities of states computed non-selfconsistently with PBE KS orbitals).

Another example for a challenge for ML-DFA or a $\Delta$-ML model development are models that should be simultaneously accurate for strong chemisorption and weak dispersion forces for chemistry on transition-metal surfaces. Part of the challenge is that the reference data typically consist of experimental binding and reaction energies only. A lack of quantum-chemistry or experimental references for transition-metal surface chemistry potential energy surfaces or even only for bound adsorbate geometries currently precludes the training of general, {\it e.g.}\ ML-interatomic potential-based $\Delta$-ML models to improve upon DFT for these chemistries. Semi-empirical models with relatively few fitting degrees of freedom for semi-local XC and non-local correlation are thus currently more suitable to be optimized with respect to the very limited amounts of transition-metal surface chemistry reference data.

\begin{figure}[b!]
\centering
\includegraphics[width=\textwidth,keepaspectratio=true]{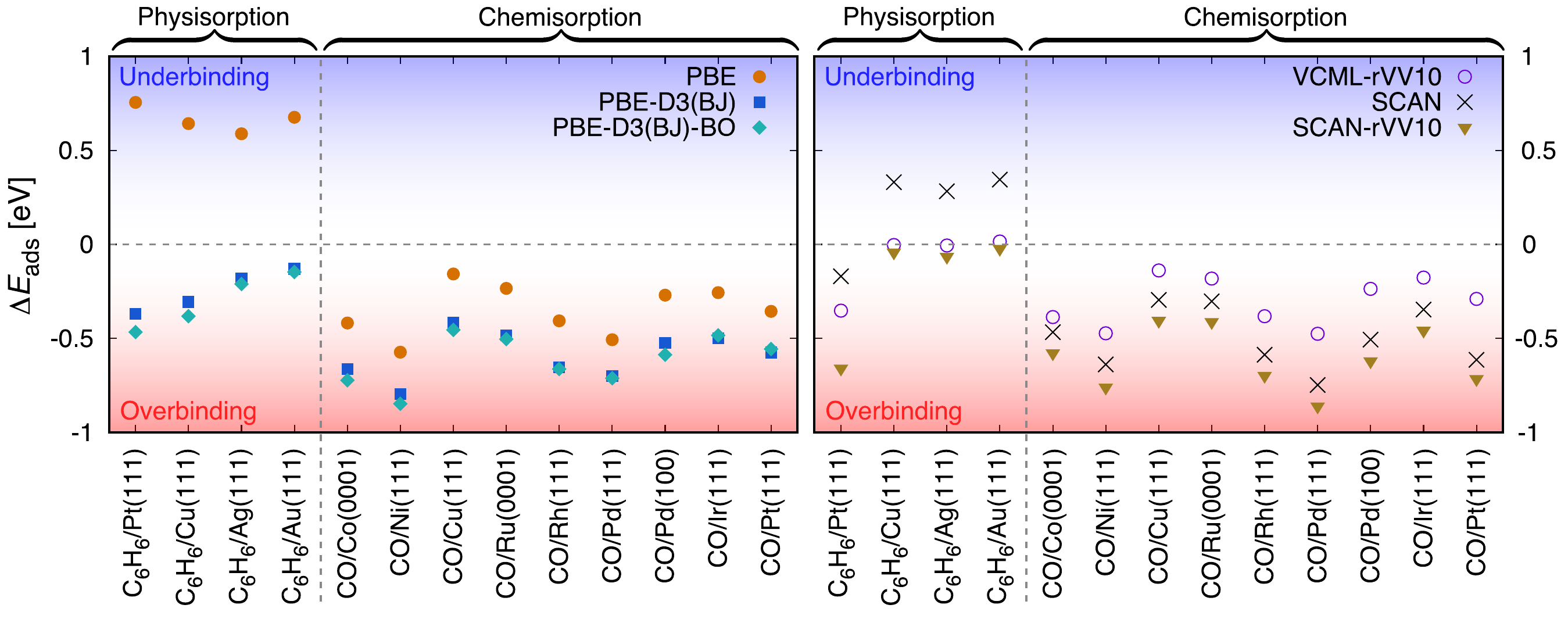}
\caption{\label{fig:Adsorption} Deviations $\Delta{}E_{\rm ads}$ of DFT transition-metal surface binding energy predictions from experiments (experimental results and PBE-based zero-point energy contributions from Ref.~\citeonl{Wellendorff2015}). PBE, VCML-rVV10, SCAN, and SCAN-rVV10 results from Ref.~\citeonl{Trepte2022}. Shown as an example for physisorption is the binding of benzene (C$_6$H$_6$) and for chemisorption the binding of carbon monoxide (CO).}
\end{figure}

A more fundamental challenge for DFAs is a general trend towards predicting too strong chemisorption, and supplementing such a DFA with attractive dispersion forces will generally only worsen the overbinding behavior for chemisorption.\cite{ADS41} Figure~\ref{fig:Adsorption} shows the accuracy of a few DFAs for predicting chemi- and physisorption energies on transition-metal surfaces. Compared are the deviations from experiments for the PBE\cite{PBE} functional, PBE supplemented with dispersive forces fields of D3-type\cite{GrimmeD3} with Becke-Johnson\cite{bjdamping} damping\cite{grimme_effect_2011} and the Bayesian optimization-tuned PBE-D3(BJ)-BO,\cite{Proppe2019} SCAN\cite{SCAN}, SCAN supplemented with a nonlocal VV10-type\cite{Vydrov2010} density functional (SCAN-rVV10),\cite{PhysRevX.6.041005} and the bulk and surface chemistry-optimized VCML-rVV10.\cite{Trepte2022} With binding of C$_6$H$_6$ as representative example for physisorption and of CO for chemisorption, PBE consistently underbinds the former and overbinds the latter. Adding dispersion energetics to PBE in form of the PBE-D3(BJ) and PBE-D3(BJ)-BO functionals improves qualitatively the binding energy of C$_6$H$_6$ on Cu, Ag, and Au, which at the PBE-level is barely bound by $\lesssim 0.05$~eV per C$_6$H$_6$. However, both PBE-D3(BJ) and PBE-D3(BJ)-BO overestimate physisorption energies of C$_6$H$_6$ in comparison to experiments, and the overbinding of PBE for chemisorption is significantly increased. PBE-D3(BJ) and PBE-D3(BJ)-BO perform very similarly for the surface chemistry examples, PBE-D3(BJ)-BO (over-)binding slightly more strongly by $\sim{}0.03$~eV on average.

Simulations using the SCAN functional predict moderate overbinding of C$_6$H$_6$ on Pt and reduced underbinding on Cu, Ag, and Au. CO is more strongly overbound than is predicted with the PBE functional. The SCAN-rVV10 functional was developed by fitting rVV10 parameters against the S66 dataset. Physisorption of C$_6$H$_6$ on Cu, Ag, and Au is described well by this functional, but for C$_6$H$_6$ on Pt and for CO chemisorption, the overbinding of SCAN is worsened. In the VCML-rVV10 approach, not only rVV10 was reparameterized, but also the semi-local part of the DFA. Physisorption on Cu, Ag, and Au is described accurately, while all other considered systems are overbound. The simultaneous reparameterization of semi-local XC allowed to suppress overbinding for chemisorption in comparison to the other considered functionals, in particular the functionals with vdW terms. However, overbinding of CO is only suppressed by $\sim{}0.03$~eV on average compared to PBE. Overbinding of chemisorbed species could not be suppressed further without breaking analytical constraints or sacrificing physisorption energetics or description of bulk lattice constants.\cite{Trepte2022} This example demonstrates difficulties in optimizing for competing interaction types with such simpler types of DFA models.

Quite generally, the description of molecular chemistry, correlations due to strongly localized states and metallic screening and delocalized states within the same electronic structure approach is an inherently difficult problem (and a very important one, given the technological importance of interfaces between oxides and metals and their defect chemistries and heterogeneous catalysis). Since imposing analytical constraints and higher levels of theory in semi-local DFT and hybrid DFT turn out to worsen the description of metallic screening, a successful XC approximation is likely going to be a highly nonlocal one. Given this likely complexity of an accurate XC functional, here is thus an opportunity for ML approaches in developing such needed XC approximations.

A major challenge for extending accurate ML DFA and $\Delta$-ML models to extended systems and metallic phases in particular is that wave-function methods will generally not be applicable. While there is progress in advanced electronic structure methods for solids,\cite{Olsen2019,Kent2020,PhysRevLett.130.036401} extending approaches to also yield accurate densities important for XC model development, once sufficient accuracy will be reached for energetics leading to a range of training data, will be a major challenge in itself.

\begin{figure}[b!]
\centering
\includegraphics[width=\textwidth,keepaspectratio=true]{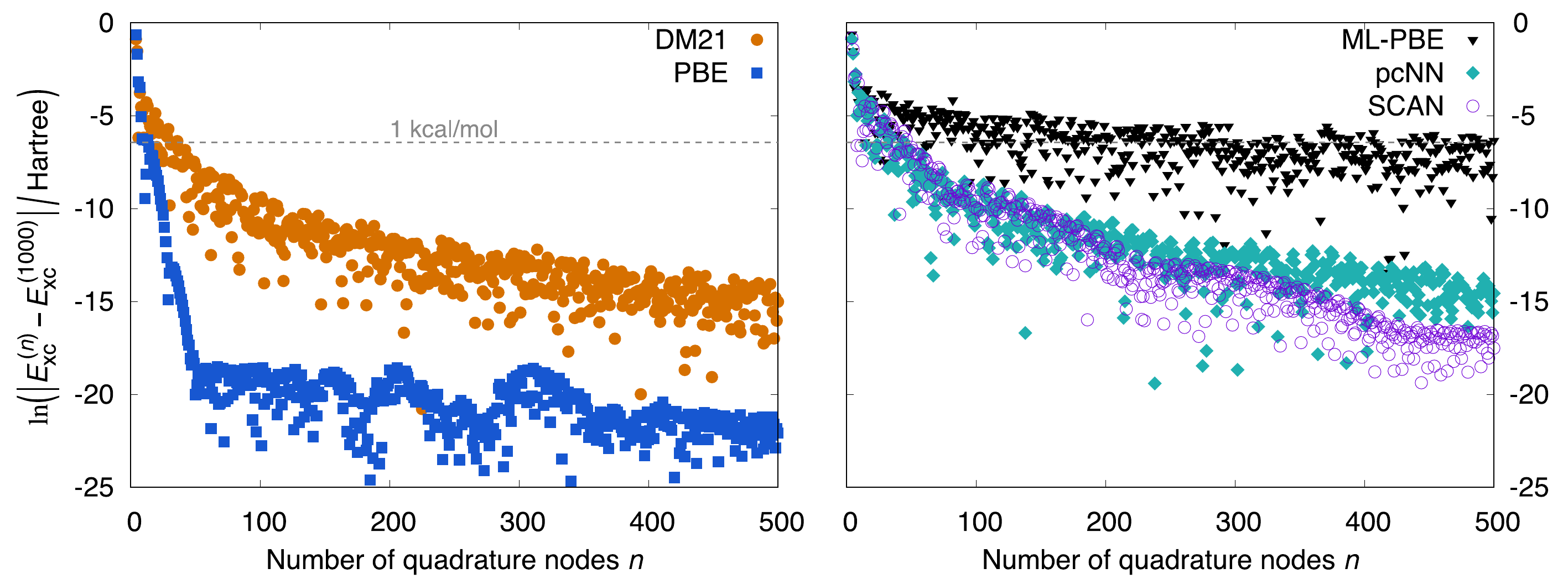}
\caption{\label{fig:Convergence} Convergence of computed XC energies for atomic O at fixed HF orbitals as a function of number of radial grid quadrature nodes. The absolute difference with respect to the results at 1000 nodes is shown.}
\end{figure}

Even if sufficient amounts of accurate training and validation data to build ML XC models are available, a remaining question will be how practical these models are for use in simulations in terms of their computational cost and numerical stability. A stability issue with DFA approximations (empirical and non-empirical) of increasing complexity can be an increased sensitivity with respect to basis set choices. Lehtola and Marques\cite{LehtolaJCP2022} have shown that several XC functionals are numerically sensitive to the number of radial quadrature nodes for evaluating atomic XC energies, requiring very fine integration grids with a large number of nodes for accurate, converged results in some cases.

Here, we perform a similar convergence test with respect to radial integration grids for the O atom (see Figure~\ref{fig:Convergence}). Similar to the findings of Lehtola and Marques,\cite{LehtolaJCP2022} the more complex meta-GGA SCAN is found to be more sensitive to the integration grid than the GGA PBE, where PBE converges much faster with respect to the number of quadrature nodes than the other, more complex DFAs tested here. To test the corresponding stability of ML approaches, we test three representative approaches: pcNN, which was trained as an enhancement over SCAN showing improved but similar performance for atomic structural and energetic metrics, DM21, as a more complex neural net-based ML DFA than pcNN, and the gradient boosting $\Delta$-ML approach ML-PBE.\cite{WangXGBoost2022} ML-PBE not being differentiable, the numerical sensitivity is worst for this approach. Even at 500 quadrature nodes, the O XC energy has not converged below a residual error of 1~kcal/mol. While this $\Delta$-ML approach would not suffer from selfconsistent cycle instabilities, as selfconsistent iterations are only performed at the PBE-level, and while there could be error cancellation in $\Delta$-ML total energy differences, this shows that predictions with this approach show a sensitivity to the basis set, that could limit its accuracy unless trained for a specific basis set or integration grid. Despite its similarity to SCAN, pcNN shows a stronger numerical sensitivity than the former with an order of magnitude higher residual error in integrated O XC energies at 500 grid points. The more complex DM21 shows a similar XC energy convergence rate as pcNN. This shows that ML DFA and $\Delta$-ML models irrespective of moderate or large model complexity can display numerical sensitivity which could render their practical use more challenging than that of simpler DFAs. Regularization techniques, relevant in particular for stable functional derivatives of ML DFAs,\cite{LiPRL2021} will play a crucial role in reducing numerical problems of these ML models.

\section{\sffamily \large METHODS}

The presented ML DFA benchmarks in the previous section were computed with the following tools. XC energies for hydrogenic ions were evaluated on their analytical nonrelativistic densities through numerical quadrature via SciPy.\cite{scipy} XC energies for the O atom were computed on HF ground-state orbitals with spherically averaged densities. HF orbitals were computed using the OPIUM code.\cite{YangPRB2018Hybrid} Spherically averaged charge, kinetic, EXX, and screened EXX energy densities were computed with these HF orbitals. B-splines (as implemented in SciPy) were used to interpolate from the employed OPIUM grid with 1335 radial grid points to sparser grids for testing numerical XC energy integration stability. For this grid convergence testing, 3 to 1000 Gauss-Legendre quadrature nodes $t$ and corresponding weights $w$ were transformed as $t^\prime = 1/(1-t) - 1/2$ and $w^\prime = w/(1-t)^2$ to evaluate the semi-infinite radial integrals. 

Charge density and KS orbitals for bulk Si were computed using the Quantum Espresso plane-wave DFT code,\cite{GiannozziJPCM2009} with a plane-wave cut-off of 600~eV and 16$\times$16$\times$16 $k$-points sampling the first Brillouin zone. Si ionic cores were described by an SG15 optimized norm-conserving Vanderbilt pseudopotential.\cite{PhysRevB.88.085117,SCHLIPF201536} Spin density and KS orbitals of bulk Fe were computed with the Vienna Ab initio Simulation Package\cite{vasp} with a 500~eV plane-wave cut-off and 28$\times$28$\times$28 $k$-points sampling the first Brillouin zone. Fe ionic cores were described by a projector-augmented wave\cite{BloechlPAW} (PAW) potential.\cite{KresseJoubert99} An additional PBE calculation of bulk Fe was performed using Quantum Espresso and a higher plane-wave cut-off of 600~eV and a norm-conserving SG15 pseudopotential, to avoid having to compute PAW augmentation terms in the evaluation of DM21 and DM21mu. DM21 and DM21mu ML XC functionals were evaluated using the C++ interface provided at \citeonl{deepmindgithub}. The corresponding bandstructure calculations were performed non-selfconsistently with charge density and KS kinetic, EXX, and long-range screened EXX energy densities evaluated with PBE KS orbitals. The zero-wave vector divergence of the Coulomb potential for the EXX computations was treated using the method due to Gygi and Baldereschi.\cite{PhysRevB.34.4405} Band dispersion plots were interpolated with the Wannier90 code.\cite{Wannier90}

Surface adsorption energies were computed with VASP at plane-wave cut-offs of 1000~eV using PAW potentials to represent ionic cores of transition-metal and adsorbate atoms. The surfaces were modeled as slabs of four layers of transition-metal atoms separated by at least 15~{\AA} of vacuum from their periodic images. The two bottom layers were fixed at their bulk positions. Forces on the two top layers and adsorbates were relaxed with residual forces of less than $10^{-2}$~eV/\AA. Brillouin zone sampling was performed with in-plane $k$-point spacings of at most $0.018$~\AA$^{-1}$.

\section{\sffamily \large SUMMARY}

Quantum chemistry benchmarks and advances in ML approximations to electronic XC enable the development of DFAs targeting chemical accuracy for a range of molecular chemistries. With semi- and nonlocal charge and energy density inputs, ML XC models can be constructed correcting for fundamental limitations of existing DFAs. A challenge for developing chemically accurate ML models for extended systems is the computation of accurate training benchmark energies and densities for these systems. Numerical stability and transferability issues to systems outside the training and validation data are a general concern. Putting these new ML XC developments to the test by the computational chemistry research community will reveal the strengths of these methods and where further development and training data are required.

\section*{\sffamily \large ACKNOWLEDGMENTS}

Support from the U.S.\ Department of Energy, Office of Science, Office of Basic Energy Sciences, Chemical Sciences, Geosciences, and Biosciences Division, Catalysis Science Program to the SUNCAT Center for Interface Science and Catalysis is gratefully acknowledged.

\bibliography{main}

\end{document}